\documentclass[12pt]{iopart}
\usepackage{units}
\usepackage{graphicx}
\usepackage{cite}
\usepackage{amssymb}
\usepackage{textcomp}

\begin{document}
\title[Thermodynamic properties of (NH$_4$)$_2$\lbrack FeCl$_5$(H$_2$O)\rbrack]{Thermodynamic properties of the new multiferroic material (NH$_4$)$_2$[FeCl$_5$(H$_2$O)]}

\author{M Ackermann$^{1}$, D Br\"uning$^{2}$, T Lorenz$^{2}$,
P Becker$^{1}$ and L Bohat\'y$^1$}

\address{$^1$ Institut f\"ur Kristallographie, Universit\"at zu K\"oln, Greinstra\ss e 6, 50937 K\"oln, Germany}
\address{$^2$ II. Physikalisches Institut, Universit\"at zu K\"oln, Z\"ulpicher Stra\ss e 77, 50937 K\"oln, Germany}

\ead{ladislav.bohaty@uni-koeln.de}

\begin{abstract}
(NH$_4$)$_2$[FeCl$_5$(H$_2$O)], a member of the family of antiferromagnetic  $A_2$[Fe$X_5$(H$_2$O)] compounds ($X$ = halide ion, $A$ = alkali metal or ammonium ion) is classified as a new multiferroic material. We report the onset of ferroelectricity below $\simeq \unit[6.9]{K}$ within an antiferromagnetically ordered state ($T_N\simeq \unit[7.25]{K}$). The corresponding electric polarization can drastically be influenced by applying magnetic fields. Based on measurements of  pyroelectric currents, dielectric constants and magnetization we characterize the magnetoelectric, dielectric and magnetic properties of (NH$_4$)$_2$[FeCl$_5$(H$_2$O)]. Combining these data with measurements of thermal expansion, magnetostriction and specific heat, we derive detailed magnetic field versus temperature phase diagrams. Depending on the direction of the magnetic field up to three different multiferroic phases are identified, which are separated by a magnetically ordered, but non-ferroelectric phase from the paramagnetic phase. Besides these low-temperature transitions, we observe an additional phase transition at $\simeq \unit[79]{K}$, which we suspect to be of structural origin.
\end{abstract}

\submitto{\NJP}
\maketitle

\section{Introduction}

Multiferroic materials with simultaneous ferroelectric and (anti-)ferromagnetic order in the same phase have attracted considerable interest during the last decade~\cite{Kimura2003,fiebig2005revival, spaldin2005renaissance, cheong2007multiferroics}. Especially  the discovery of spin-driven ferroelectricity in magnetically frustrated systems \cite{kimura2007spiral}, as, e.g., in the transition metal oxides $RE$MnO$_3$ ($RE={\rm Tb}$, Dy)~\cite{kimura2005magnetoelectric}, Ni$_3$V$_2$O$_8$~\cite{Lawes2005}, LiCu$_2$O$_2$~\cite{Park2007}, MnWO$_4$ \cite{Heyer, Taniguchi, Arkenbout}, NaFe$X_2$O$_6$ ($X$=Si, Ge)~\cite{Jodlauk,Kim2012}, CuO~\cite{Kimura2008} or CaMn$_7$O$_{12}$~\cite{Johnson2012} (but also in non-oxide systems such as, e.g., CuCl$_2$~\cite{Seki2010} or K$_3$Fe$_5$F$_{15}$~\cite{Blinc2008}) revived the search for new multiferroic materials. Typically, these multiferroics show complex, non-collinear spin structures and a strong coupling between magnetic and ferroelectric order exists. Consequently, the spontaneous electric polarization can strongly be modified by applying external magnetic fields. Depending on the direction and strength of the magnetic field, reversal, rotation or suppression of the electric polarization may occur. Such magnetic field induced changes of the electric  polarization or, vice versa, electric field dependent magnetization changes, are not only interesting from the fundamental physical point of view, but are also interesting for potential new devices in the fields of data memory or sensor systems.

\begin{figure}[h]
\includegraphics[width=\textwidth]{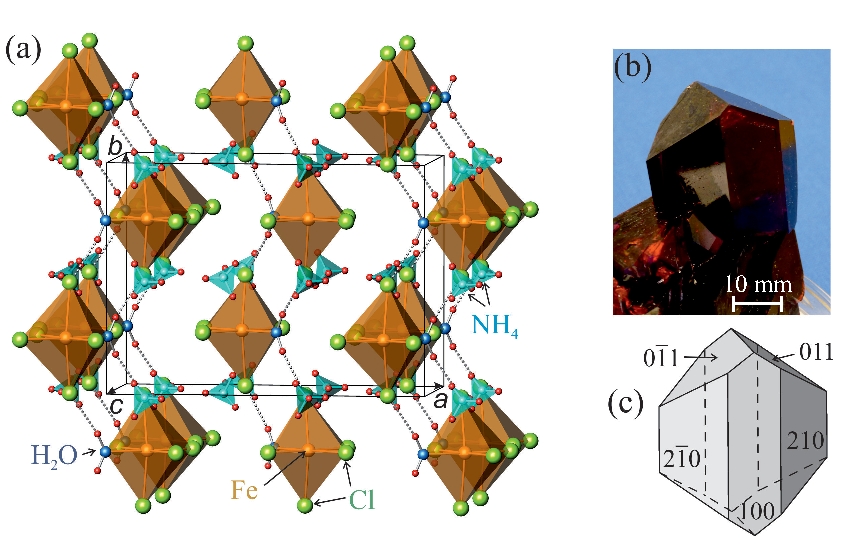}
\caption{(a) Crystal structure of (NH$_4$)$_2$[FeCl$_5$(H$_2$O)]; NH$_4$ groups are displayed by blue tetrahedra (with H atoms in red), Fe and Cl are marked by orange and green spheres, respectively, while the oxygen of H$_2$O is given by blue spheres. The unit cell is marked by solid lines and  the dashed lines indicate the hydrogen bonds between the H$_2$O and Cl ligands of adjacent [FeCl$_5$(H$_2$O)] octahedra. Structural data are taken from~\cite{Figgis1978a}. A photograph of a  grown (NH$_4$)$_2$[FeCl$_5$(H$_2$O)] crystal and the typical growth morphology with face indices are displayed in (b) and (c), repectively.}
\label{strukt}
\end{figure}

Here, we report the discovery and the basic characterization of the new multiferroic material ammonium pentachloroaquaferrate(III), (NH$_4$)$_2$[FeCl$_5$(H$_2$O)]. It belongs to the family of erythrosiderite-type compounds $A_2$[Fe$X_5$(H$_2$O)], where $A$ stands for an alkali metal or ammonium ion and $X$ for a halide ion. As for most members of this family, the room-temperature crystal structure of (NH$_4$)$_2$[FeCl$_5$(H$_2$O)] is orthorhombic with space group $\textit{Pnma}$~\cite{bellanca1948} and lattice constants (at room temperature) $a=\unit[13.706(2)]{\r{A}}$, $b=\unit[9.924(1)]{\r{A}}$, $c=\unit[7.024(1)]{\r{A}}$~\cite{Figgis1978a}. The structure consists of isolated (NH$_4$)$^+$ units and isolated complex groups [FeCl$_5$(H$_2$O)]$^{2-}$ of sixfold octahedrally coordinated iron(III), see figure~\ref{strukt}. The unit cell contains eight symmetrically equivalent (NH$_4$)$^+$ groups and four [FeCl$_5$(H$_2$O)]$^{2-}$ octahedra. The Fe--O bonds of the octahedra are  approximately lying parallel to the $ac$ plane with alternating angles of about $\pm 41^{\circ}$ relative to the $\bi{a}$ axis. This results in a herringbone-like arrangement of the Fe--O bonds along $\bi{a}$. Besides ionic bonds between the structural building blocks, there is a pronounced H-bonding \mbox{(via O--H--Cl)} between neighbouring [FeCl$_5$(H$_2$O)]$^{2-}$ octahedra along the $\bi{b}$ axis, which further stabilizes the crystal structure. These H-bonded octahedra form zig-zag chains along $\bi{b}$ and along these chains the Fe-O bonds of adjacent octahedra are oriented mutually antiparallel to each other \cite{bellanca1948, Figgis1978a, Lackova}.

The magnetic ordering phenomena occurring in the $A_2$[Fe$X_5$(H$_2$O)] series have been subject of various investigations and all members studied so far have been identified as antiferromagnets~\cite{McElearney1978a,Carlin1985a,Carlin,Connor,Gabas1995a,Campo2008a}. In this context, a N\'{e}el temperature $T_{N}\simeq \unit[7.25]{K}$ of the title compound has been derived from measurements of the magnetic susceptibility~\cite{McElearney1978a}. However, there are some clear differences between (NH$_4$)$_2$[FeCl$_5$(H$_2$O)] and the corresponding alkali compounds with $A= {\rm K}$, Rb. Susceptibility data of the alkali compounds identify the $\bi{a}$ axis as the magnetic easy axis~\cite{McElearney1978a,Carlin,Connor}, in agreement with the antiferromagnetic collinear spin structure determined via neutron scattering~\cite{Gabas1995a, Campo2008a}. In contrast, no easy axis could be derived from the susceptibility data of (NH$_4$)$_2$[FeCl$_5$(H$_2$O)], and, moreover, heat capacity measurements revealed a second phase transition at $\unit[6.87]{K}$, while only single transitions are observed in the alkali compounds~\cite{McElearney1978a,Puertolas1985}. These observations led to the suggestion of a canted antiferromagnetic spin structure with some kind of spin rearrangement at $\unit[6.87]{K}$. Later on, there was some debate about the proposed spin canting and the occurrence of a structural phase transition around \unit[240]{K} was suggested~\cite{Palacio1980a,Partiti1985a}. However, no evidence for such a structural transition has been found in M\"ossbauer spectroscopy and powder X-ray diffraction experiments within the relevant temperature range~\cite{Partiti1988a}. The M\"ossbauer spectra also led to the suggestion of a more complex spin arrangement in (NH$_4$)$_2$[FeCl$_5$(H$_2$O)], because the interpretation of the spectra below $T_N$ was inconsistent with a simple canted spin structure~\cite{Partiti1988a}. To our knowledge, an unambiguous determination of the magnetic structure of  (NH$_4$)$_2$[FeCl$_5$(H$_2$O)] is still missing.

In the present work, by means of dielectric, pyroelectric and magnetoelectric measurements, (NH$_4$)$_2$[FeCl$_5$(H$_2$O)] is established as a new multiferroic material. Our data agree with the suggested N\'{e}el ordering at $\simeq \unit[7.25]{K}$ followed by a spin-reorientation at $\simeq \unit[6.9]{K}$ and we find, that this lower transition is accompanied by a pronounced spontaneous electric polarization. The polarization and magnetization data are complemented by measurements of 
thermal expansion, magnetostriction and heat capacity in order to derive detailed magnetic field versus temperature phase diagrams for different directions of the magnetic field.

\section{Experiments}
\label{exp}
Large single crystals of (NH$_{4}$)$_{2}$[FeCl$_{5}$(H$_{2}$O)] were grown from aqueous solution with a surplus of HCl and a non-stoichiometric ratio of the educts \mbox{FeCl$_{3}$ : NH$_{4}$Cl = 2~:~1} at \unit[311]{K} by slow controlled evaporation of the solvent. After typical growth periods of 6--8 weeks optically clear, red single crystals with dimensions up to about \mbox{40$\times$30$\times\unit[20]{mm^{3}}$} and well-developed flat morphological faces $\lbrace$100$\rbrace$, $\lbrace$210$\rbrace$, $\lbrace$101$\rbrace$ and $\lbrace$011$\rbrace$ were obtained, see figure \ref{strukt}\,(b). Using the morphological faces as reference planes, oriented samples with faces perpendicular to the $\bi{a}$, $\bi{b}$ or $\bi{c}$ axis were prepared. The magnetic susceptibility and the specific heat were measured on  plate-like samples of typical thicknesses of $\sim$\,$\unit[1]{mm}$ and surfaces of $\sim$\,$\unit[5]{mm^2}$, while the dielectric investigations were performed on plates of larger surfaces ($\sim$\,$\unit[30]{mm^2}$), which were metallized by silver electrodes. For the measurements of thermal expansion and magnetostriction we used cube-shaped samples of $\sim$\,\unit[15--150]{mm$^3$} with edges along $\bi{a}$, $\bi{b}$ and $\bi{c}$. 

X-ray powder diffraction measurements were performed in the temperature range from 5 to \unit[300]{K} on powdered single crystals using a diffractometer Siemens D-5000 equipped with a home-built $^4$He-flow cryostat. Si was added to the powder samples as an internal standard. Peak positions and intensities were determined from the raw data by full-profile fitting, the patterns were indexed with the program TREOR and lattice constants were refined by least squares refinement (all implemented in the WinXPOW suite \cite{winxpow}). The magnetization measurements were performed with a commercial SQUID magnetometer (MPMS, Quantum Design) in the temperature range from 2 to \unit[300]{K} in magnetic fields up to \unit[7]{T}, while the dielectric measurements were done in the same temperature range in a cryostat equipped with a \unit[14]{T} magnet. The relative dielectric constants $\epsilon_{i}^{r}$ ($i = a, b, c$) were calculated from the capacitances, measured by a capacitance bridge (Andeen-Hagerling 2500A) at fixed frequency of \unit[1]{kHz}. The temperature and magnetic field dependence of the electric polarization was determined by time-integrating the pyroelectric and magnetoelectric currents, respectively, which were measured by an electrometer (Keithley 6517). In order to ensure a single-domain state, a static electric poling field of at least \unit[200]{V/mm} was applied well above the transition temperature. After cooling the crystal to base temperature, the poling field was removed and the pyroelectric current was recorded during the subsequent  heating process with constant heating rates between \unit[0.5]{K/min} and \unit[3]{K/min}. For the magnetoelectric current measurements, the same field-cooled poling process of the crystals was performed and the magnetoelectric currents were registered while sweeping the magnetic field with a rate of \unit[0.4]{T/min} at fixed temperatures. In contrast to the pyroelelectric current measurements, here the electric poling field was present during the whole measurement process. In all cases the electric polarization could be completely inverted by reversing the electric poling field.

Magnetostriction and low-temperature thermal expansion were measured on a home-built capacitance dilatometer in the temperature range between \unit[250]{mK} and \unit[10]{K} using a $^3$He evaporator system (Heliox-VL, Oxford) in a \unit[15]{T} magnetcryostat. The magnetic field was applied either along the $\bi{a}$, $\bi{b}$, or $\bi{c}$ axes and in each case the longitudinal length change $\Delta L_i(T,\bi{B}) || \bi{B}$ was calculated from the measured capacitance changes, either as a function of continuously varying $T$ or $\bi{B}$ with rates of $\pm 0.05$ to $\pm \unit[0.1]{K/min}$ or $\pm \unit[0.1]{T/min}$, respectively. In addition, the thermal expansion in zero magnetic field was measured up to \unit[150]{K} on another home-built dilatometer and between 130 and \unit[273]{K} on a commercial inductive dilatometer (Perkin Elmer TMA7, sweep rate $\pm$ \unit[1]{K/min}). The thermal expansion coefficients $\alpha_i = 1/L_i^0 \cdot \partial  \Delta L_i/\partial T$ were obtained by numerically calculating the temperature derivatives of the length changes $\Delta L_i$, where $L_i^0$ denotes the respective sample length along the measured axis $i=\bi{a}$, $\bi{b}$, or $\bi{c}$. The specific heat was measured between 2 and \unit[300]{K} in a commercial calorimeter (PPMS, Quantum Design) using a thermal relaxation method.

\section{Results and discussion}

\begin{figure}[t]
\includegraphics[width=\textwidth]{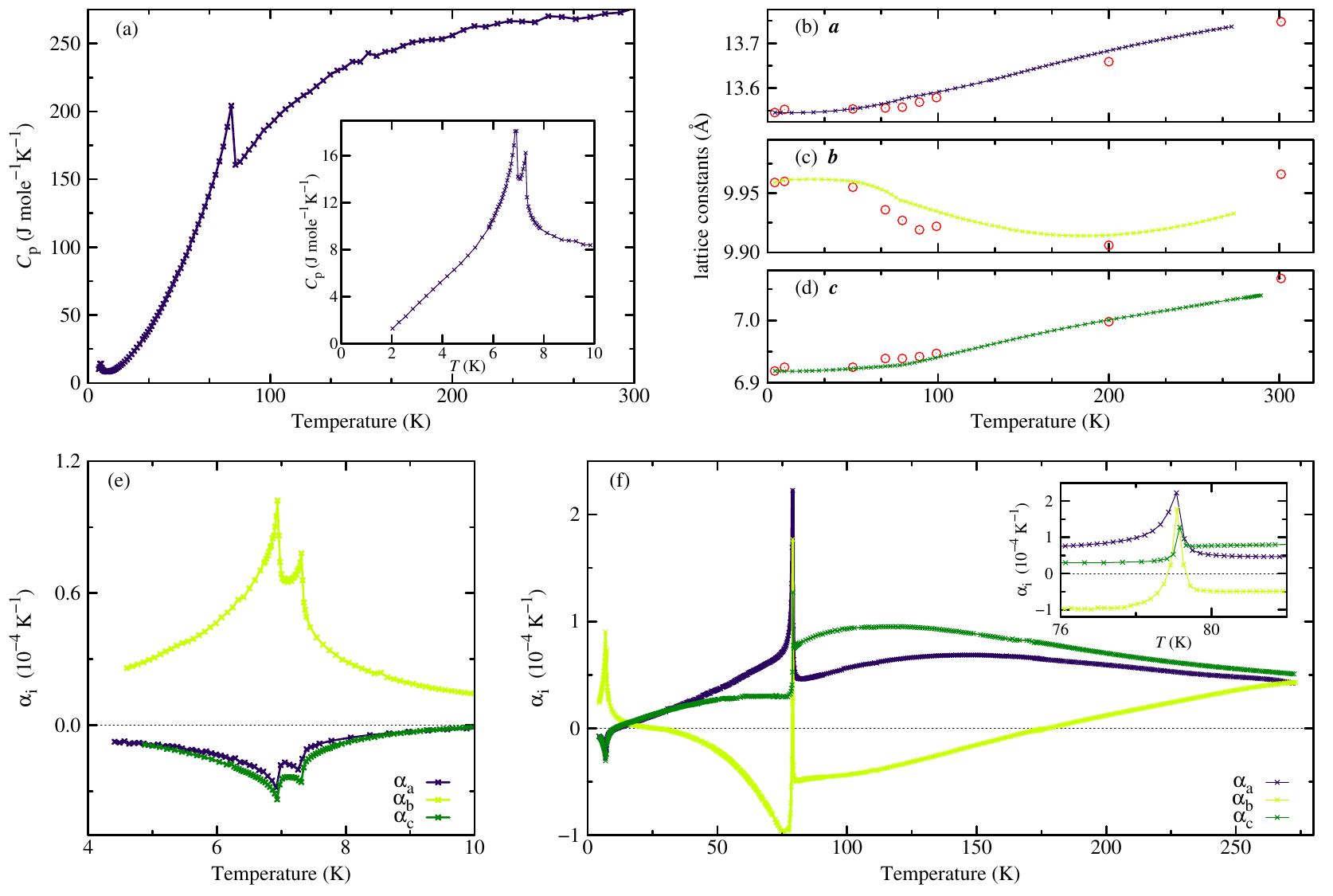}
\caption{(a): Heat capacity of (NH$_4$)$_2$[FeCl$_5$(H$_2$O)] in a wide temperature range;  the inset is a magnification of the low-temperature region. (b)-(d): Temperature dependencies of the lattice constants $\bi{a}$, $\bi{b}$, and $\bi{c}$ obtained by X-ray powder diffraction measurements (circles) in comparison to the respective macroscopic length changes (lines) measured by capacitance dilatometry. (e) and (f): Thermal expansion coefficients $\alpha_i$ along all three axes $\bi{a}$, $\bi{b}$ and $\bi{c}$ below \unit[10]{K} and over the entire temperature range, where the inset shows an expanded view around the transition at $\simeq \unit[79]{K}$.}
\label{alpha}
\end{figure}

The heat capacity measurement of (NH$_4$)$_2$[FeCl$_5$(H$_2$O)], displayed in figure \ref{alpha}\,(a),  confirms the occurrence of the phase transitions at $\simeq \unit[6.9]{K}$ and $\simeq \unit[7.3]{K}$ already reported in literature~\cite{McElearney1978a}. In addition, our data reveal a third phase transition at $T_{\rm HT} \simeq \unit[79]{K}$. These three phase transitions were also found in the measurements of all three thermal expansion coefficients $\alpha_{i}$, see figures~\ref{alpha}\,(e) and~(f). However, neither the specific heat data nor the $\alpha_{i}(T)$ curves give any hint to a phase transition around \unit[240]{K}, where the occurrence of a structural phase transition had been supposed for (NH$_4$)$_2$[FeCl$_5$(H$_2$O)]~\cite{Partiti1985a}.
The fact that no anomalies were found in the magnetic susceptibility measurements around $T_{\rm HT}$ (see below, Fig.~\ref{chi}\,(e)) led to the conjecture that the transition around \unit[79]{K} might be of structural origin. Therefore we performed temperature-dependent X-ray powder diffraction measurements, but surprisingly, the patterns  above and below $T_{\rm HT}$ are identical within the experimental uncertainty. Particularly, no peak splittings or violations of systematic extinctions of the space group $\textit{Pnma}$ of the high-temperature phase could be found below $T_{\rm HT}$.  In Figures~\ref{alpha}\,(b)--(d) the temperature-dependent lattice constants obtained from X-ray diffraction are compared with the corresponding results obtained from the measured macroscopic length changes via capacitance dilatometry. There is an overall agreement between both methods over the entire temperature range, see e.g. the anomalous negative thermal expansion of the $\bi{b}$ axis. The shape of the specific-heat anomaly at $T_{\rm HT}$ suggests a second-order phase transition, but the thermal-expansion curves, which have been studied with much higher density around $T_{\rm HT}$, clearly show rather sharp peaks of positive signs, which are present for all three directions, see inset of figure \ref{alpha}\,(f). Integrating over these peaks reveals an almost discontinuous length change $\Delta L/L^0$ of all three directions $\bi{a}$, $\bi{b}$ and $\bi{c}$ of $\simeq 0.5 - 1.5 \cdot 10^{-3}$, indicating that the transition at $T_{\rm HT}$ is of first order. Considering all this information, we suspect that the structural changes of (NH$_4$)$_2$[FeCl$_5$(H$_2$O)] at $T_{\rm HT}$ probably only affect the hydrogen atoms of the crystal structure. Consequently, the room-temperature crystal structure~\cite{Figgis1978a, Lackova} still can serve as a good approximation at lower temperatures, at least concerning the positions of the non-hydrogen atoms. Because our data do not indicate a lowering of the orthorhombic symmetry to the monoclinic (or even triclinic) system, the (NH$_4$)$_2$[FeCl$_5$(H$_2$O)] crystals will be described as being orthorhombic also below $T_{\rm HT}$ in the following, although this assumption should be treated with some caution.

The results of the temperature dependent magnetic susceptibility measurements of (NH$_4$)$_2$[FeCl$_5$(H$_2$O)] for magnetic fields up to \unit[7]{T} (applied along the $\bi{a}$, $\bi{b}$ and $\bi{c}$ axis) are summarized in figures \ref{chi}\,(a)--(c), respectively. The low-field curves are consistent with previous results~\cite{McElearney1978a} and signal magnetic ordering at $T_{\rm N} \simeq \unit[7.3]{K}$. While $\chi_{a}$ and $\chi_{c}$ show well-defined kinks at $T_N$ and subsequently decrease with decreasing temperature, $\chi_{b}(T)$ hardly changes below $T_N$, indicating that the spins are lying in the $ac$ plane. The fact that $\chi_{a}$ and $\chi_{c}$ of (NH$_4$)$_2$[FeCl$_5$(H$_2$O)] behave very similar and none of them approaches zero for  $T\rightarrow \unit[0]{K}$ implies, that there is no magnetic easy axis within the $ac$ plane. This is further supported by additional low-field susceptibility measurements (not shown) with magnetic fields along other directions within the $ac$ plane. Thus, the spins of (NH$_4$)$_2$[FeCl$_5$(H$_2$O)] show an XY anisotropy with the $ac$ plane as magnetic easy plane. As mentioned above, there have been different speculations about possible spin orientations within the $ac$ plane earlier~\cite{McElearney1978a,Palacio1980a,Partiti1985a,Partiti1988a}, but no definite magnetic-structure determination has been reported until now.

\begin{figure}[t]
\includegraphics[width=\textwidth]{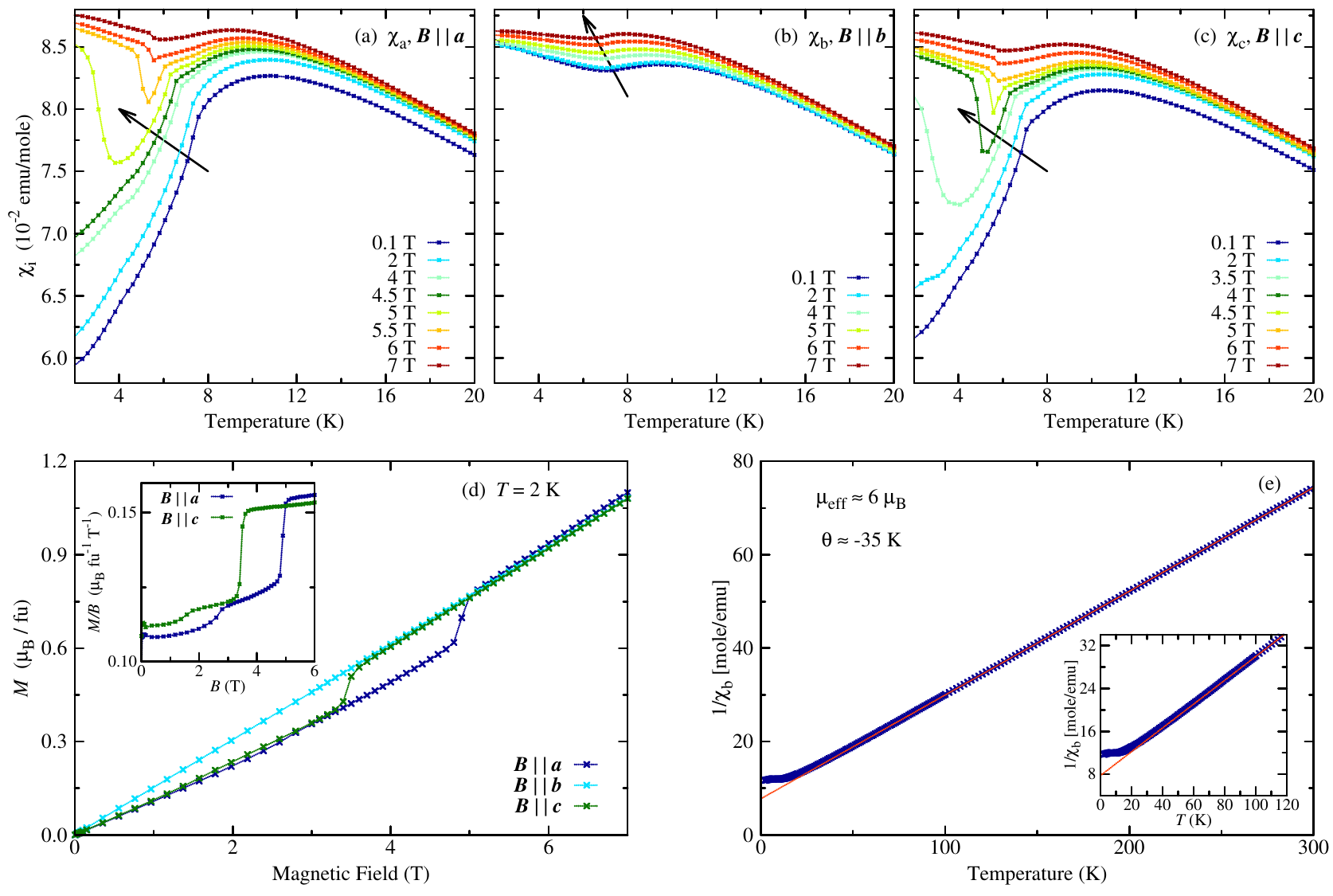}
\caption{(a)--(c): Magnetic susceptibility $\chi_{a}$, $\chi_{b}$, and $\chi_{c}$ of (NH$_4$)$_2$[FeCl$_5$(H$_2$O)] for different magnetic fields. The arrows mark the direction of increasing field strength. (d): Magnetization at \unit[2]{K} as functions of the magnetic field applied parallel $\bi{a}$, $\bi{b}$, and $\bi{c}$; the inset shows $M/B$ for $\bi{B}||\bi{a}$ and $\bi{B}||\bi{c}$. (e): Temperature dependence of the inverse susceptibilty $1/\chi_{\mathrm{b}}$ (the inset shows a magnification of the low-temperature range).}
\label{chi}
\end{figure}

Concerning the magnetic anisotropy, (NH$_4$)$_2$[FeCl$_5$(H$_2$O)] differs from the analogous potassium and rubidium compounds $A_2$[FeCl$_5$(H$_2$O)] ($A={\rm K}$, Rb).  According to low-field susceptibility measurements~\cite{McElearney1978a,Carlin, Connor}, $\chi_{a}$ of both alkali metal compounds approaches zero for $T \rightarrow \unit[0]{K}$, while $\chi_{b}$ and $\chi_{c}$ hardly change, identifying $\bi{a}$ as the magnetic easy axis. This is confirmed by magnetic-structure determinations of K$_2$[FeCl$_5$(D$_2$O)] and Rb$_2$[FeCl$_5$(D$_2$O)] by neutron scattering investigations, which in addition revealed the Shubnikov symmetry $Pn'm'a'$ for these compounds~\cite{Gabas1995a,Campo2008a}. These neutron studies also allowed to determine five relevant exchange couplings $J_{1}-J_{5}$, which are all found to be antiferromagnetic. There are three stronger ones ($J_{1}-J_{3}$) which enforce an antiferromagnetic ordering between the spins lying in neighbouring $ac$ planes at $y=0.25$ and $y=0.75$. However, the spins lying in the same $ac$ plane are ordered ferromagnetically, despite of the antiferromagnetic, but weaker interactions $J_{4}$ and $J_{5}$ within these planes \cite{Gabas1995a,Campo2008a}. Thus, there exists a certain degree of magnetic frustration in the $A_2$[FeCl$_5$(H$_2$O)] structure, which might be the reason for the more complex spin arrangement in the compound with $A={\rm (NH_4)}$, compared to those with $A={\rm K}$, Rb.     

As shown in figure~\ref{chi}\,(a) and (c), towards larger fields the decrease for $T<T_N$ of both, $\chi_{a}(T)$ and $\chi_{c}(T)$, systematically vanishes and finally above about \unit[6]{T} the principal behavior of $\chi_{i}(T)$ is almost identical for all three field directions. This suggests that for either $\bi{B}|| \bi{a}$ and  $\bi{B}|| \bi{c}$, there are spin-flop transitions, with a change of  spin orientation from lying within the easy $ac$ plane to lying within the plane that is perpendicular to the respective magnetic field direction. These spin-flop transitions can also be seen in the low-temperature measurements of the magnetization as a function of the magnetic field, see figure~\ref{chi}\,(d). The spin-flop fields at $T=\unit[2]{K}$ are $\simeq \unit[5]{T}$ and $\simeq \unit[3.5]{T}$ for $\bi{B}|| \bi{a}$ and  $\bi{B}|| \bi{c}$, respectively. Besides the spin-flop transitions, the magnetization curves display additional anomalies for both field directions, which are better resolved in a representation $M/B$ versus $B$, see inset of figure~\ref{chi}\,(d). Corresponding anomalies can be seen in the temperature dependent susceptibility data in figure \ref{chi}\,(a) and (c) for $B=\unit[4]{T}$ and \unit[2]{T}, respectively. As will be seen below, both types of transitions are related to structural changes and as well to reorientations of the electric polarization. The fact that both transition fields are larger for $\bi{B}|| \bi{a}$ than for $\bi{B}|| \bi{c}$ illustrates, that the magnetic properties of (NH$_4$)$_2$[FeCl$_5$(H$_2$O)] are not completely isotropic within the easy $ac$ plane. This is supported by the slight, but systematic differences between the $\chi_{a}(T)$ and $\chi_{c}(T)$ curves in the intermediate field range, see figures~\ref{chi}\,(a) and (c).

Figure \ref{chi}\,(e) displays the inverse susceptibility $1/\chi_{b}$ for a magnetic field of \unit[0.1]{T} applied along $\bi{b}$, which follows a Curie-Weiss behaviour from above about \unit[40]{K} up to room temperature. Here, we only show the data of $\chi_{b}$, because almost identical values are obtained for $\chi_{c}$ in this high-temperature range, while $\chi_a$ is slightly larger ($\simeq 1-2~\%$). Almost isotropic high-temperature susceptibilities may be expected for a $3d^5$ high-spin configuration of the Fe$^{3+}$ ions, as was already stated in a previous work~\cite{McElearney1978a}. Linear fits to the data of  $1/\chi_{i}$ for $T> \unit[100]{K}$ yield a negative Weiss temperature $\theta \simeq \unit[-35]{K}$ and an effective magnetic moment $\mu_{\mathrm{eff}}\simeq \unit[6]{\mu_{B}}$. The negative Weiss temperature signals a net antiferromagnetic exchange interaction and the effective magnetic moment is close to the expected value $\mu_{\mathrm{eff}}=g\mu_{B}\sqrt{S(S+1)}=\unit[5.92]{\mu_{B}}$ of Fe$^{3+}$ with $S=5/2$. 

\begin{figure}[t]
\centering
\includegraphics[width=\textwidth]{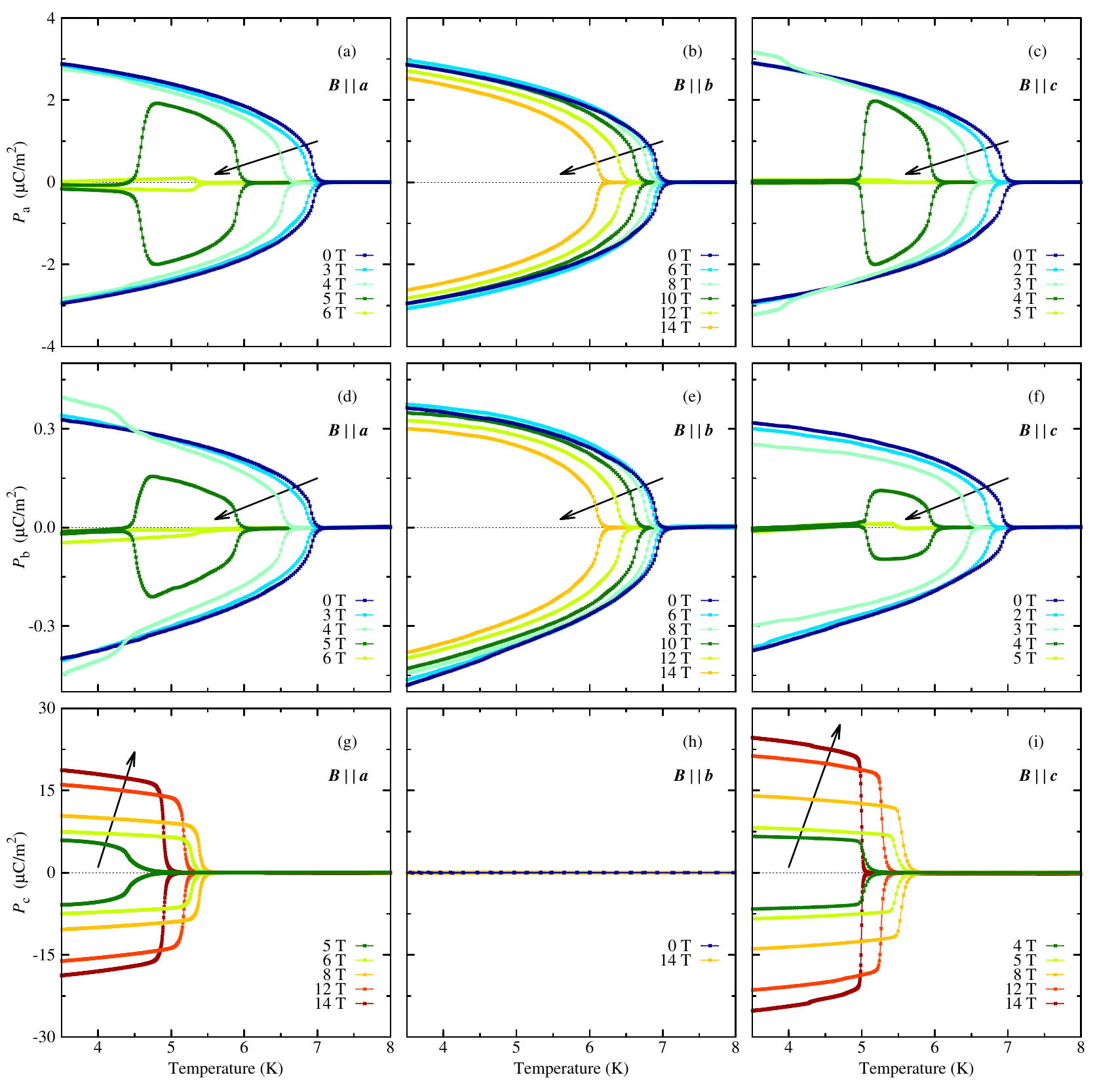}
\caption{Temperature dependencies of the spontaneous electric polarization $P_a$, $P_b$ and $P_c$ (top to bottom) of (NH$_4$)$_2$[FeCl$_5$(H$_2$O)] for magnetic fields applied parallel to the $\bi{a}$, $\bi{b}$ or $\bi{c}$ axis (left to right). In all cases, single-domain phases were obtained by cooling the samples with an applied electric field of at least \unit[200]{V/mm}. Besides the electric polarization is completely invertible by changing the sign of the electric field. Both directions of the electric polarization are shown, which were obtained by integrating the measured pyroelectric currents during the heating process after the poling field was removed at base temperature. The directions of increasing magnetic fields are marked by arrows.}
\label{pol1}
\end{figure}

Figure~\ref{pol1} summarizes the results of the temperature-dependent measurements of the electric polarization $P_{a}$, $P_{b}$ and $P_{c}$ in magnetic fields applied along $\bi{a}$, $\bi{b}$ and $\bi{c}$. The positive and negative values of $\bi{P}$, depending on the direction of the electric poling field, agree with each other within the experimental uncertainties, indicating that the electric polarization can be completely inverted.  In zero magnetic field, (NH$_4$)$_2$[FeCl$_5$(H$_2$O)] becomes ferroelectric below $T_{\mathrm{FE}}\simeq \unit[6.9]{K}$ and develops a spontaneous electric polarization, with the components $P_{a}\simeq \unit[3]{\mu C/m^{2}}$, $P_{b}\simeq \unit[0.3]{\mu C/m^{2}}$ and $P_{c}\simeq 0$ at \unit[3]{K}. Thus, the spontaneous electric polarization is lying in the $ab$ plane with an angle relative to the $\bi{a}$ axis of \mbox{$\sphericalangle(\bi{P},\bi{a})\simeq$\,$7^\circ$}. This marked deviation of $\bi{P}$ from the crystallographic $\bi{a}$ axis indicates that the symmetry of this ferroelectric phase is most probably lower than orthorhombic. Assuming a point group symmetry  $2/m$~$2/m$~$2/m$ for the prototypic phase, the probable crystallographic symmetry of the considered ferroelectric phase would be $m$ (with cell setting unique axis $\bi{c}$) or even triclinic, $1$.

\begin{figure}[t]
\includegraphics[width=\textwidth]{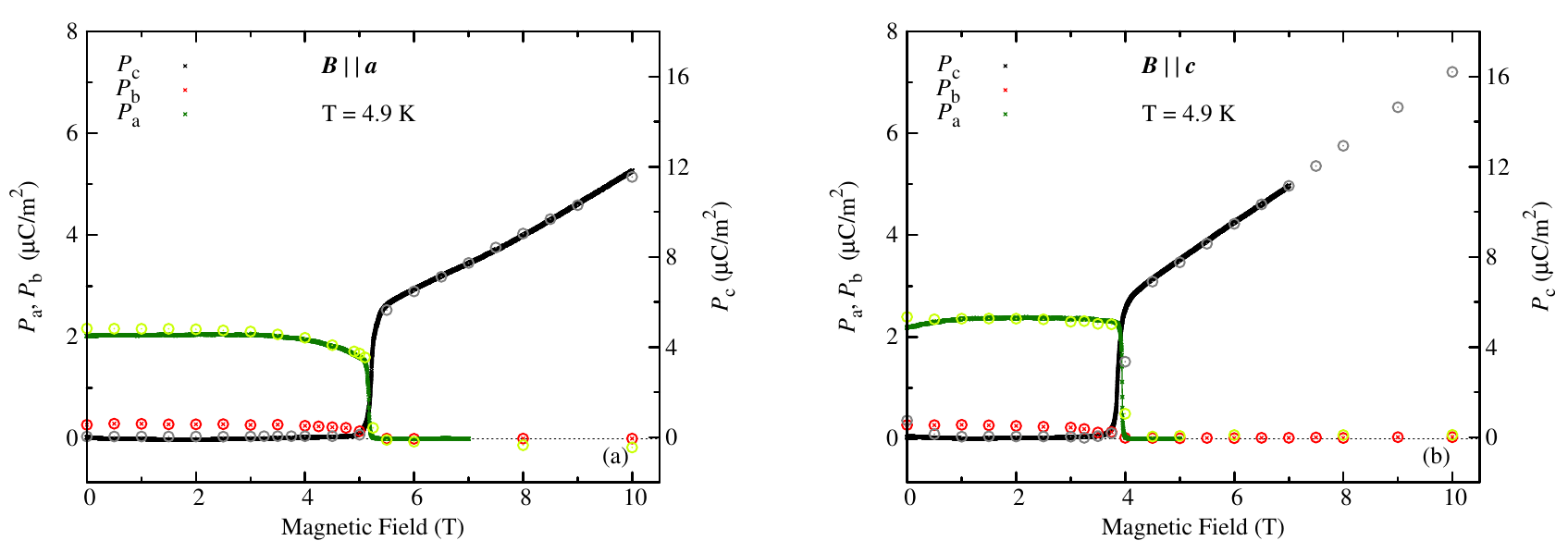}
\caption{Representative magnetic field dependencies of the electric polarization $P_a$, $P_b$ and $P_c$ for magnetic fields applied along $\bi{a}$ (left) and $\bi{c}$ (right) at $T=\unit[4.9]{K}$. Open symbols indicate data taken from the temperature-dependent polarization measurements shown in figure~\ref{pol1} and the solid lines indicate $P_{a}(\bi{B})$ and $P_{c}(\bi{B})$ obtained by integrating the magnetoelectric currents measured during magnetic field sweeps. The latter method is not possible for $P_b(\bi{B})$ due to its small absolute value. Note the different scales used for $P_{i}$ ($i=a,b$) and for $P_c$.}
\label{pol2}
\end{figure}

A magnetic field applied parallel $\bi{b}$ causes a weak systematic reduction of the transition temperature \mbox{($T_{\mathrm{FE}}^{14\mathrm{T}}\simeq \unit[6]{K}$)} and also slightly reduces the magnitude of the electric polarization, but leaves its orientation almost unchanged. In contrast, magnetic fields applied either along $\bi{a}$ or $\bi{c}$ have a much stronger influence on the electric polarization. Most obviously, for both directions a field of about \unit[4--6]{T}  causes a complete suppression of both components $P_a$ and $P_b$, and simultaneously induces a finite polarization $\bi{P}||\bi{c}$, which continuously increases with further increasing magnetic field. In figures~\ref{pol2}\,(a) and (b), these magnetic field induced abrupt reorientations of $\bi{P}$ for $\bi{B}||\bi{a}$ and $\bi{B}||\bi{c}$, respectively, are shown exemplarily for $T=\unit[4.9]{K}$ by plotting the components  $P_a$, $P_b$ and $P_c$ as functions of the magnetic field. The open symbols are taken from the temperature-dependent polarization data of figure~\ref{pol1}, while the solid lines were obtained by integrating the magnetoelectric currents during a magnetic field sweep as described in section~\ref{exp}. For $P_a(\bi{B})$ and $P_c(\bi{B})$, the results of both methods agree to each other, while the polarization $P_b(\bi{B})$ is too small to induce large enough magnetoelectric currents, even if the magnetic field sweep rate is increased to \unit[1]{T/min}. A comparison of the $\bi{P}(\bi{B})$ data with the magnetization curves $\bi{M}(\bi{B})$ discussed above reveals, that the magnetic field dependent reorientations of the electric polarization coincide with the spin-flop transitions for both field directions  $\bi{B}||\bi{a}$ and $\bi{B}||\bi{c}$. In addition, the data of figure~\ref{pol2} show, that the electric polarization $P_c$ above the critical field grows linearly with increasing magnetic field. At $B=\unit[14]{T}$, $P_{c}\simeq \unit[19]{\mu C/m^{2}}$ or $\simeq \unit[25]{\mu C/m^{2}}$ is reached for $\bi{B} || \bi{a}$ or $\bi{B} || \bi{c}$, respectively, i.e., the absolute value of the polarization is enhanced by a factor of about 6 to 8 compared to its zero-field value. If the field-induced increase of the electric polarization above the transition is eliminated by linearly extrapolating $P_c(B>B_{\mathrm{crit}})$ back to the critical field, the absolute value of the polarization, e.g.\ at \unit[4.9]{K}, changes from $|\bi{P}|\simeq P_a \simeq \unit[2]{\mu C/m^{2}}$ below $B_{\mathrm{crit}}$ to $|\bi{P}|= P_c \simeq \unit[5.2]{\mu C/m^{2}}$ for $\bi{B}||\bi{a}$ and to $|\bi{P}|= P_c \simeq \unit[5.6]{\mu C/m^{2}}$ for $\bi{B}||\bi{c}$, respectively, above $B_{\mathrm{crit}}$.

As already discussed above, besides the spin-flop transition the magnetization data for both field directions, $\bi{B}||\bi{a}$ and $\bi{B}||\bi{c}$, signal another phase transition, which is located about \unit[2]{T} below the spin-flop and is seen as small step-like increases of $M/B$, see inset of figure~\ref{chi}\,(d). These additional transitions also cause small changes in the electric polarization, which can be best seen by considering the turquoise curves in figures~\ref{pol1}\,(c) and (d). For $B=\unit[3]{T}$ along $\bi{c}$ the polarization $P_a(T)$ shows a distinct increase around \unit[4]{K}, while the polarization $P_b(T)$ has an additional increase around \unit[4.5]{K} for $B=\unit[4]{T}$ along $\bi{a}$ . 

\begin{figure}[t]
\includegraphics[width=\textwidth]{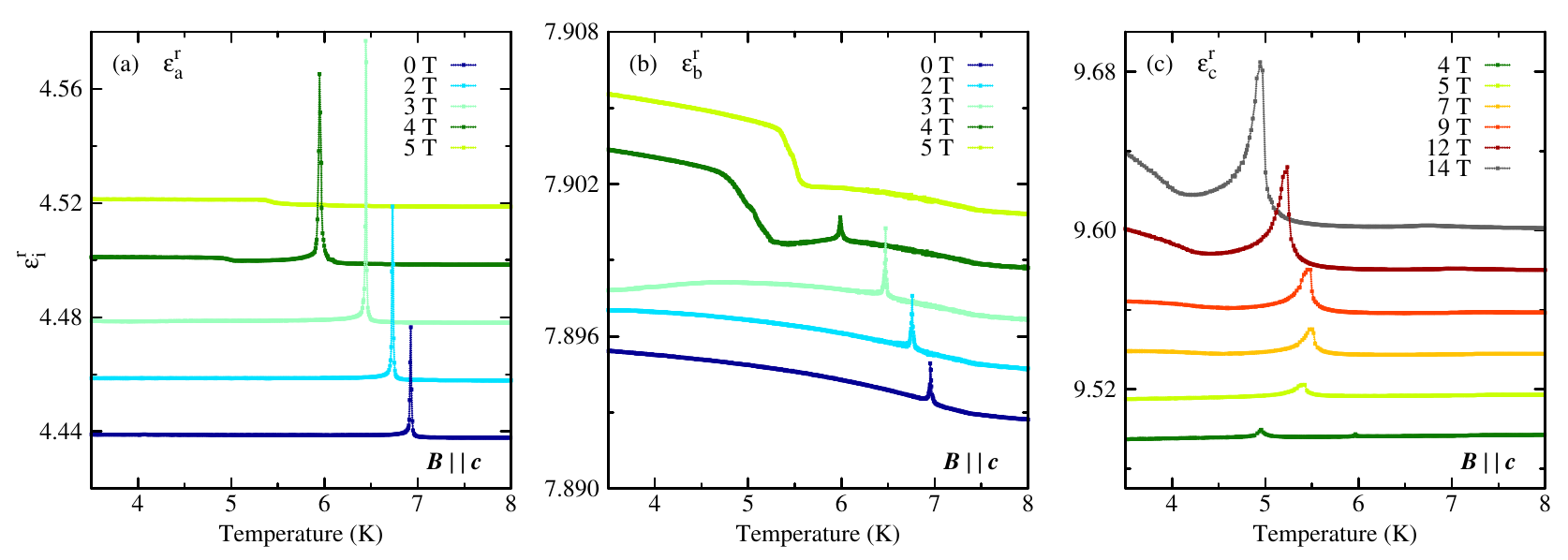}
\caption{Representative temperature dependencies of the longitudinal components of the dielectric tensor $\epsilon^{r}_{i}$
($i=a, b, c$ from left to right) for different magnetic fields applied parallel $\bi{c}$. For clarity, with increasing field strength the curves are shifted with respect to each other by constant offsets of 0.02 in (a) \& (c) and by 0.002 in (b).}
\label{DK}
\end{figure}

In summary, the magnetic field dependent (re-)orientations of the electric polarization of  (NH$_4$)$_2$[FeCl$_5$(H$_2$O)] can be described as follows: In zero magnetic field the spontaneous polarization is lying in the $ab$ plane and almost pointing along the $\bi{a}$ axis \mbox{($\sphericalangle(\bi{P},\bi{a})\simeq$\,$7^\circ$)}. A magnetic field applied along $\bi{a}$ first (at $\simeq \unit[4]{T}$) causes a slight reorientation of the polarization in the $ab$ plane towards the $\bi{b}$ axis \mbox{($\sphericalangle(\bi{P},\bi{a})\sim$\,$9^\circ$)}. At a larger field of about \unit[5]{T} the polarization then completely rotates by 90$^\circ$ from the $ab$ plane to the $\bi{c}$ axis. For a magnetic field along $\bi{c}$, the first transition (at $\simeq \unit[3]{T}$) causes a reorientation of the spontaneous polarization in the opposite sense \mbox{($\sphericalangle(\bi{P},\bi{a})\sim$\,$5^\circ$)}, but is also followed by an analogous 90$^\circ$ change of the polarization to the $\bi{c}$ axis, occurring at about \unit[4.5]{T}. For both field directions, the first transitions are accompanied by weak anomalies in the magnetization, while the 90$^\circ$ reorientations of the electric polarization from the $ab$ plane to the $\bi{c}$ axis coincide with spin-flop transitions from the magnetic easy $ac$ plane towards the plane perpendicular to the respective magnetic field direction. Interestingly, for both magnetic field directions the 90$^\circ$ reorientation of the electric polarization is accompanied by an abrupt, significant enhancement of its absolute value by about a factor of 2.5 and on further increasing the magnetic field the polarization linearly grows with field. Loosely speaking, a magnetic field $\bi{B}||\bi{c}$ has a "stronger effect" than a magnetic field $\bi{B}||\bi{a}$: The high-field slope $\partial P_c/\partial B_c \simeq \unit[1.7]{A/V} $ is larger than $\partial P_a/\partial B_a \simeq  \unit[1.2]{A/V}$ and in order to induce the phase transition smaller field strengths are needed for $\bi{B}||\bi{c}$ than for $\bi{B}||\bi{a}$.

The transitions to the ferroelectric phases, discussed above, also cause distinct anomalies in the temperature and magnetic field dependencies of the corresponding longitudinal components $\epsilon_{i}^{r}$ of the dielectric tensor. This is shown exemplarily for $\bi{B}||\bi{c}$ in figure~\ref{DK}, which displays representative measurements of the temperature dependencies of $\epsilon_{i}^{r}$ ($i=a, b, c$). Below \unit[5]{T}, $\epsilon_{a}^{r}(T)$ and $\epsilon_{b}^{r}(T)$ show spiky anomalies in the temperature range between 6 and \unit[7]{K}, which essentially vanish for larger fields, while such spiky anomalies of growing intensity occur in $\epsilon_{c}^{r}(T)$ in this higher field range. From these results, together with the polarization data above, it can be concluded that these spikes signal the corresponding transition temperatures to the ferroelectric phases with a spontaneous polarization lying either in the $ab$ plane or being oriented along the $\bi{c}$ axis. There are additional anomalies, e.g. the step-like increase in $\epsilon_{b}^{r}(T)$ for 4 and \unit[5]{T}, which are more difficult to interpret.    

\begin{figure}[t]
\includegraphics[width=\textwidth]{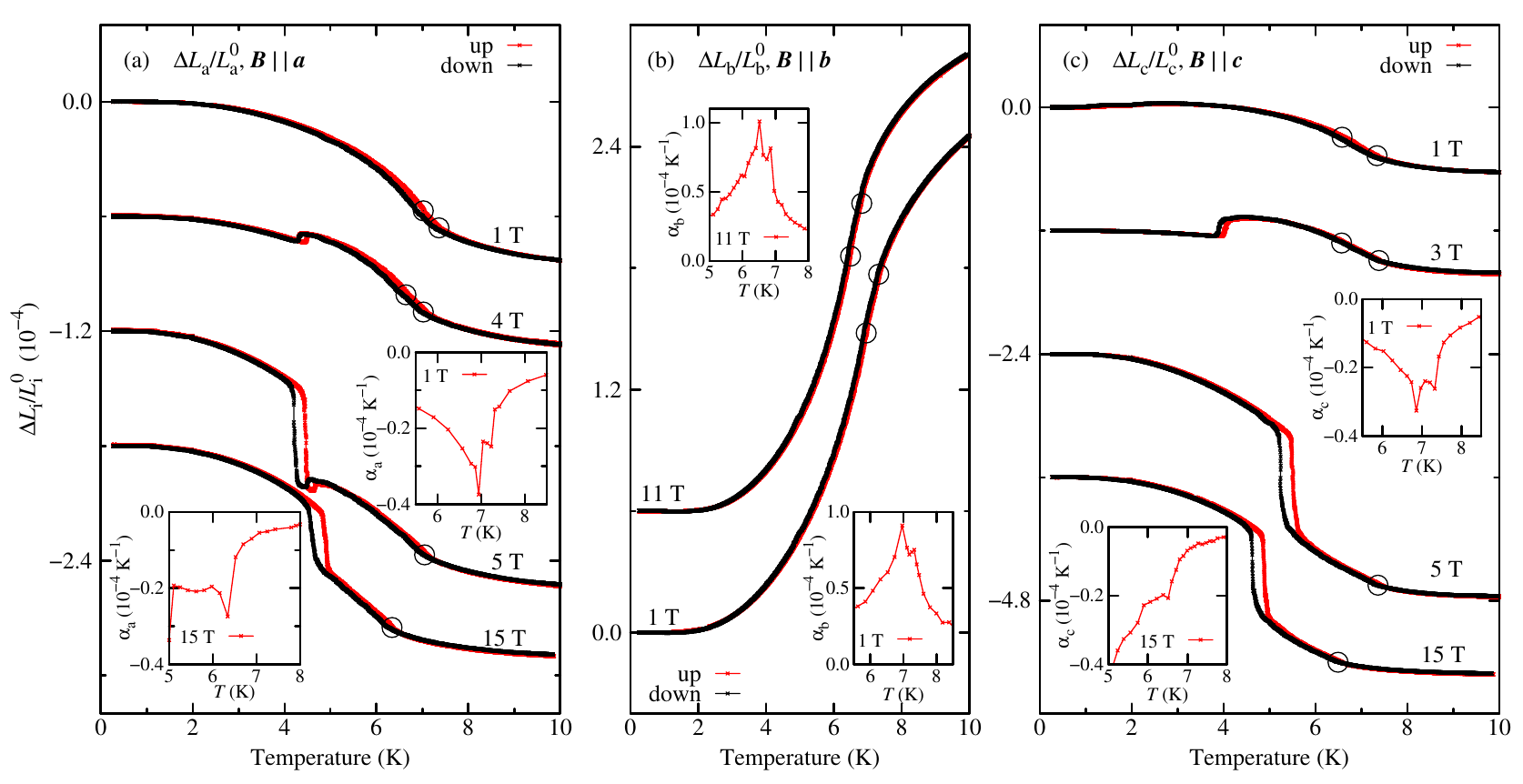}
\caption{Thermal expansion of (NH$_4$)$_2$[FeCl$_5$(H$_2$O)] along the $\bi{a}$, $\bi{b}$ and $\bi{c}$ direction for representative magnetic fields applied parallel $\bi{a}$, $\bi{b}$ and $\bi{c}$, respectively. For clarity, with increasing field strengths the curves are shifted with respect to each other by constant offsets of $0.6\cdot 10^{-4}$ in (a) \& (b) and by $1.2\cdot 10^{-4}$ in (c). Measurements with increasing and decreasing temperature are plotted as red and black symbols, respectively. The weak anomalies due to the continuous transitions at $T_{\rm FE}$ and $T_{\rm N}$ (see the zero-field data in figure~\ref{alpha}) are hardly visible in the $\Delta L_i/L_i^0$ curves, but can be better seen in the thermal expansion coefficients $\alpha_i$ shown in the insets. The corresponding transition temperatures are marked by open circles on the $\Delta L_{i}/L_{i}^0$ curves.} 
\label{hystTA}
\end{figure}

The discussion above has mainly focused on the dielectric and magnetic properties of (NH$_4$)$_2$[FeCl$_5$(H$_2$O)], which are obviously most sensitive to transitions where either the polarization or the magnetization changes. In order to map out the temperature versus magnetic field phase diagrams of this multiferroic material, it is more convenient to study thermodynamic properties, which are sensitive to both types of phase transitions, such as the specific heat or the thermal expansion, see figure~\ref{alpha}. In the present case, we have performed detailed studies of the thermal expansion and of the magnetostriction for all three magnetic field directions $\bi{a}$, $\bi{b}$ and $\bi{c}$. As will be seen, (NH$_4$)$_2$[FeCl$_5$(H$_2$O)] shows various first-order phase transitions with rather large, almost discontinuous changes of the sample dimensions. Thus, in the following we will present the relative length changes $\Delta L_i(T,\bi{B})/L_i^0$ with $\bi{B}||L_i$ for $i=a, b$ and $c$. Figure~\ref{hystTA} displays a representative selection of $\Delta L_i(T,\bi{B})/L_i^0$ curves measured as a function of increasing or decreasing temperature at constant magnetic fields. For $\bi{B}||\bi{b}$, there is only little change. The two transitions occurring at $T_{\rm FE} \simeq \unit[6.9]{K}$ and $T_{\rm N} \simeq \unit[7.3]{K}$ in zero field are just continuously shifted towards lower temperatures with increasing field. As shown in the insets of  figure \ref{hystTA}\,(b), the corresponding anomalies are resolved in the thermal expansion coefficient $\alpha_b$, but are only faintly visible in $\Delta L_b(T,\bi{B})/L_b^0$. Thus, these transitions are marked by open circles in the main panel of this figure. Additional anomalies occur for $\bi{B}|| \bi{a}$ and for $\bi{B}|| \bi{c}$. For both field directions, there is a sharp, almost discontinuous decrease of $\Delta L_i(T,\bi{B})/L_i^0$ on decreasing the temperature around \unit[4]{K} in the field range of \unit[3--4]{T}, indicating that the corresponding phase transition is of first order. Moreover, there is a weak temperature hysteresis, but its width partly arises also from the finite temperature sweep rate (see section \ref{exp}).
With a further increase of the field $\bi{B}|| \bi{a}$ or $\bi{B}|| \bi{c}$ above about \unit[5]{T}, this discontinuous anomaly disappears again and also the continuous anomaly at the transition temperature $T_{\rm FE}$ vanishes, but another first-order phase transition occurs. This latter transition is characterized by a more pronounced hysteresis and a discontinuous length change $\Delta L_i(T,\bi{B})/L_i^0$, which is significantly larger  and of the opposite sign than that of the discontinuous transition occurring below \unit[5]{T}.

\begin{figure}[t]
\includegraphics[width=\textwidth]{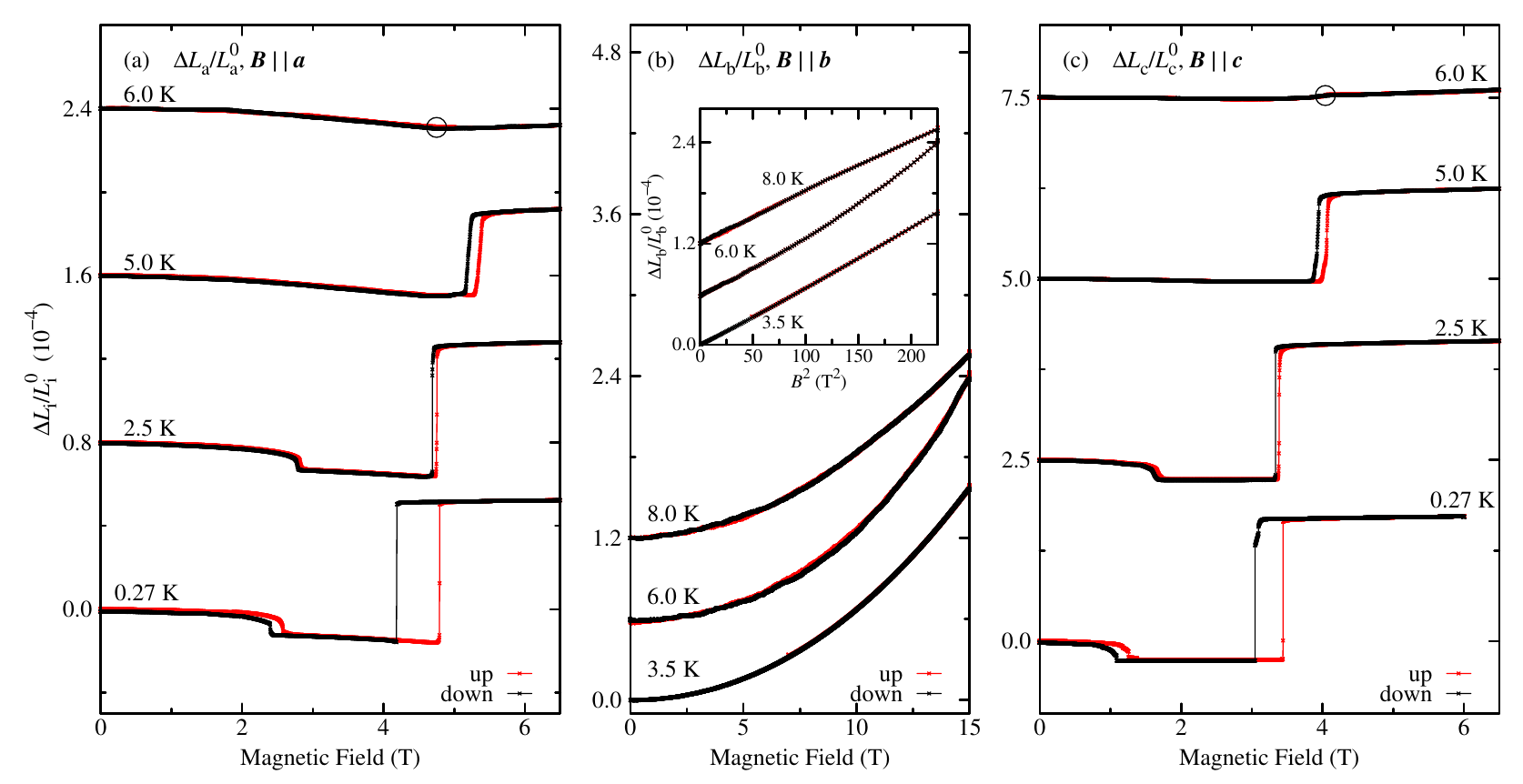}
\caption{Magnetostriction of (NH$_4$)$_2$[FeCl$_5$(H$_2$O)] along the $\bi{a}$, $\bi{b}$ or $\bi{c}$ direction for magnetic fields applied parallel $\bi{a}$, $\bi{b}$ and $\bi{c}$, respectively. For clarity,  with increasing temperature the curves are shifted with respect to each other by constant offsets of $0.8\cdot 10^{-4}$, $0.6\cdot 10^{-4}$ and $2.5\cdot 10^{-4}$ in (a), (b) and (c), respectively.  Measurements with increasing and decreasing magnetic field are plotted as red and black symbols, respectively. The open circles on the $\Delta L_{i}(\bi{B}, T=6\,\rm{K})/L_{i}^0$ curves in (a) and (c) mark continuous phase transitions, which can better be seen in the corresponding derivatives with respect to $B$ (not shown). The inset in (b) shows $\Delta L_{b}/L_{b}^0$ versus $B^2$.}
\label{hystMS}
\end{figure}

Figure \ref{hystMS} summarizes representative magnetostriction measurements along the  $\bi{a}$, $\bi{b}$ or $\bi{c}$ direction for magnetic fields applied parallel $\bi{a}$, $\bi{b}$ and $\bi{c}$, respectively. Again, the data for $\bi{B}|| \bi{a}$ or $\bi{B}|| \bi{c}$ are rather similar. At the lowest temperature ($T=\unit[0.27]{K}$) there are two hysteretic first-order phase transitions with (almost) discontinuous length changes $\Delta L_{i}/L_{i}^0$ of opposite signs for both directions $\bi{a}$ and $\bi{c}$. Both transitions are also present at \unit[2.5]{K}, but the hysteresis widths are considerably smaller. At an even higher temperature of \unit[5]{K}, the transitions at the lower critical fields have vanished, while those at the larger fields are still present and their hysteresis widths are wider again. Finally at  \unit[6]{K}, the discontinuous length changes have vanished. Instead there are kinks in $\Delta L_{i}/L_{i}^0$ around \unit[4--5]{T}, which can be better seen in the corresponding derivatives with respect to $B_i$ (not shown). These anomalies signal second-order phase transitions and are marked by open circles on the $\Delta L_{i}/L_{i}^0$ curves. 
Note, that although most of the data have been studied up to a maximum field of \unit[15]{T}, the field scales of figures \ref{hystMS}\,(a) and (c) have been limited to \unit[6.5]{T}, because there are no further anomalies in the higher field range. Figure \ref{hystMS}\,(b) displays characteristic longitudinal magnetostriction curves for $\bi{B}|| \bi{b}$. As can be inferred from the inset, in the paramagnetic/non-ferroelectric phase,  i.e. above $T_{\rm N}\simeq \unit[7.3]{K}$, as well as in the multiferroic phase, i.e. well below $T_{\rm FE}\simeq \unit[6.9]{K}$, there is a quadratic magnetostriction  ($\Delta L_{b}/L_{b}^0\propto B^2$) in the entire field range studied, which is typical for materials with a linear field dependence of the magnetization. The curves taken at \unit[6]{K} deviate from this simple quadratic field dependence above about \unit[10]{T}, which is related to the fact that the field dependent phase boundary $T_{\rm FE}(\bi{B})$ is approached.

\begin{figure}[t]
\includegraphics[width=\textwidth]{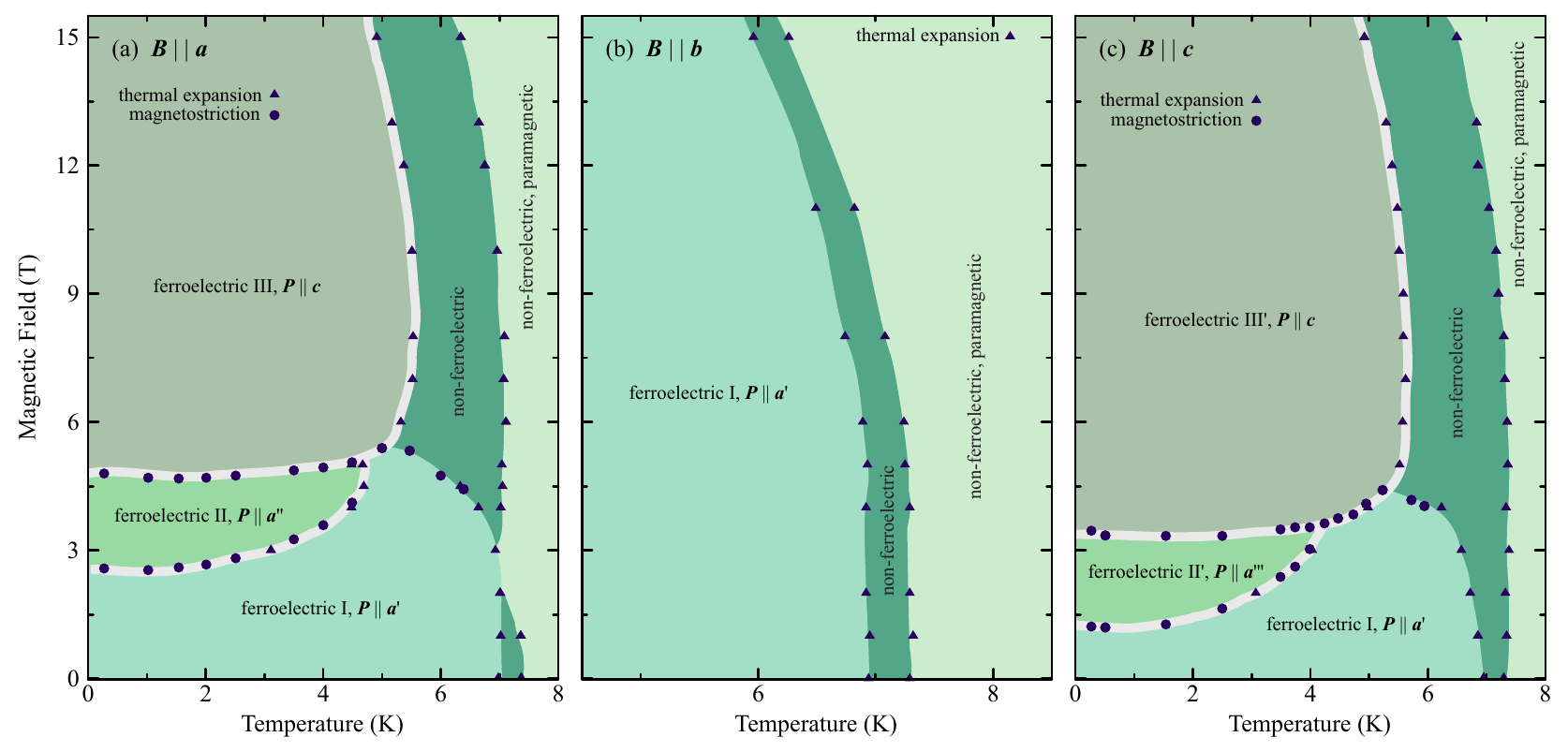}
\caption{Temperature versus magnetic field phase diagrams of (NH$_4$)$_2$[FeCl$_5$(H$_2$O)] for $\bi{B}$ parallel $\bi{a}$, $\bi{b}$ and $\bi{c}$, respectively. The phase boundaries are based on the anomalies of the relative length changes $\Delta L_{i}(T,\bi{B})/L_{i}^0$ ($i=a$, $b$ or $c$) obtained from measurements of the thermal expansion ($\opentriangle$) or the magnetostriction $(\opencircle\!\!)$ with increasing  temperature or magnetic field, respectively. Almost discontinuous (and hysteretic) first-order phase transitions are marked by bold light-grey lines. For each of the different ferroelectric phases, the direction of the spontaneous electric polarization is denoted. All the vectors $\bi{a}'$, $\bi{a}''$ and $\bi{a}'''$ lie within the $ab$ plane with small deviations from the $\bi{a}$ axis \mbox{($\sphericalangle(\bi{a},\bi{a'})\simeq $ $7^\circ$}, \mbox{$\sphericalangle(\bi{a},\bi{a''})\simeq $ $9^\circ$}, \mbox{$\sphericalangle(\bi{a},\bi{a'''})\simeq $ $5^\circ$)}. }
\label{phasendiagramm}
\end{figure}

The $B$--$T$ phase diagrams of (NH$_4$)$_2$[FeCl$_5$(H$_2$O)] for magnetic fields applied along $\bi{a}$, $\bi{b}$ or $\bi{c}$ are displayed in figure~\ref{phasendiagramm}. The critical fields and temperatures are based on the anomalies in $\Delta L_{i}(T,\bi{B})/L_{i}^0$ from the thermal expansion and magnetostriction measurements, while the dielectric and magnetic properties of the various phases are inferred from the electric polarization and magnetization data. The polarization and magnetization measurements also show anomalies for most of the detected phase transitions and their positions  agree well with the anomalies of $\Delta L_{i}(T,\bi{B})/L_{i}^0$ and also with those found in specific heat measurements, which were performed for $\bi{B}||\bi{a}$ (not shown). On temperature decrease in zero magnetic field, (NH$_4$)$_2$[FeCl$_5$(H$_2$O)] undergoes a phase transition at $T_{\rm N}\simeq$\,$\unit[7.3]{K}$ from its paramagnetic and non-ferroelectric phase to an antiferromagnetically ordered non-ferroelectric phase.  On further cooling, ferroelectricity occurs at $T_{\rm FE}\simeq$\,$\unit[6.9]{K}$. The ferroelectric phase extends down to our experimental low-temperature limit of \unit[0.25]{K}. The spontaneous polarization of this  ferroelectric phase I is oriented along the direction $\bi{a}'$ in the $ab$ plane with a small deviation from the $\bi{a}$ axis of \mbox{$\sphericalangle(\bi{a},\bi{a'})\simeq $ $7^\circ$}. The detailed magnetic structure of this phase (as well as of all the other phases) is not known, but from the magnetization data it can be concluded that the spins are lying within the $ac$ plane without a magnetic  easy axis. As mentioned above, neutron scattering data revealed a certain frustration potential for the analogous compounds $A_2$[FeCl$_5$(D$_2$O)] with $A=$~K, Rb~\cite{Gabas1995a, Campo2008a}. Assuming similar magnetic exchange interactions for (NH$_4$)$_2$[FeCl$_5$(H$_2$O)], one may speculate that small structural changes due to the substitution with \mbox{$A=$ NH$_{4}$} could further enhance this magnetic frustration and lead to the occurrence of spin spirals in the $ac$ plane.  This then  could act as the underlying mechanism of multiferroicity of the title compound via inverse Dzyaloshinsky-Moriya interaction \cite{Mostovoy2006, Katsura2005}. In this case the fact, that the spins are lying in the $ac$ plane would imply a spin current $\bi{j}_{\mathrm{s}} = \bi{S}_{i} \times \bi{S}_{j}$ along $\bi{b}$. Therefore, the observed electric polarization, pointing nearly along $\bi{a}$ (in low magnetic fields) would require a propagation vector $\bi{r}$ of the spin spiral (nearly) along $\bi{c}$ due to $\bi{P} \sim \bi{r} \times \bi{j}_s$.

As shown in figure~\ref{phasendiagramm}\,(b), there is little influence of a magnetic field $\bi{B}||\bi{b}$, which only causes a weak simultaneous decrease of both transition temperatures $T_{\rm FE}(\bi{B})$ and $T_{\rm N}(\bi{B})$. As expected for this field direction perpendicular to the magnetic easy plane, the magnetization linearly increases with field (see figure~\ref{chi}). Moreover, a weak decrease of the electric polarization can be inferred from figures~\ref{pol1}\,(b) and (e).

Magnetic fields applied either along $\bi{a}$ or $\bi{c}$ induce discontinuous transitions to other multiferroic phases. With increasing $\bi{B}||\bi{a}$ and at lowest temperatures, first a ferroelectric phase II is entered, in which the polarization tilts to $\bi{P}||\bi{a''}$ with \mbox{$\sphericalangle(\bi{a},\bi{a''})\simeq $ $9^\circ$}. At a larger field of about \unit[4.5]{T} another discontinuous transition to a ferroelectric phase III occurs, where the spontaneous polarization increases by about a factor of 2.5 and rotates to $\bi{P}||\bi{c}$. The ferroelectric phase II only occurs below about \unit[4.7]{K}, and in a small temperature window around \unit[5]{K} direct discontinuous transitions between the ferroelectric phases I and III take place. Finally, above about \unit[5.3]{K} continuous phase transitions between the ferroelectric phase I and the non-ferroelectric, antiferromagnetic phase are observed in the field range below about \unit[5]{T}, while for higher fields discontinuous transitions between the non-ferroelectric, antiferromagnetic phase and the ferroelectric phase III take place. The transitions between the ferroelectric phases I and II cause rather weak anomalies in the magnetization, whereas those to the ferroelectric phase III are accompanied by a spin-flop transition. Most of the features described for $\bi{B}||\bi{a}$ are also present for $\bi{B}||\bi{c}$, but there are some significant differences. First of all, at the transition from the ferroelectric phase I to the intermediate phase II' the polarization rotates in the opposite sense to $\bi{P}||\bi{a'''}$ with \mbox{$\sphericalangle(\bi{a},\bi{a'''})\simeq 5^\circ$}. Secondly, the transition fields are somewhat smaller for $\bi{B}||\bi{c}$ and the phase II'' only exists up to about \unit[4]{K}. Consequently, the temperature window for a direct transition between the ferroelectric phases I and III' is wider than for $\bi{B}||\bi{a}$. Finally, for $\bi{B}||\bi{c}$, in the entire field range studied here, no direct transitions from the non-ferroelectric, paramagnetic phase to any of the multiferroic phases take place, whereas between 2 and \unit[3]{T} for $\bi{B}||\bi{a}$ there seems to be a small window for a direct transition to phase I. A direct transition of this type would require two different order parameters to be involved simultaneously. A similar case has been debated recently for CuO~\cite{Giovannetti2011,Jin20012,Toledano2011}, where a direct transition from the non-ferroelectric, paramagnetic phase to the multiferroic phase had been reported originally. But more recently a sequence of two second-order phase transitions with very close critical temperatures, instead of one first-order phase transition was found in CuO~\cite{Villarreal2012, Quirion2013}. At present, we cannot exclude that this might also be the case in (NH$_4$)$_2$[FeCl$_5$(H$_2$O)], but our actual data of both, thermal expansion and specific heat do not give any hint for the presence of two phase transitions in the field range of 2 to \unit[3]{T} for $\bi{B}||\bi{a}$.    

Among all ferroelectric phases of the phase diagrams given in figure \ref{phasendiagramm} the ferroelectric phases III and III' possess the most pronounced magnetoelectric effect, with a clear linear magnetic field dependence of the polarization. From the slope \mbox{$\partial P_i/\partial B_j=\alpha_{ij}^{B}$} derived in connection with figure \ref{pol2} (see above) the linear magnetoelectric coefficients \mbox{$\alpha_{ca}^{B}\simeq \unit[1.2\cdot 10^{-6}]{A/V}$} and \mbox{$\alpha_{cc}^{B}\simeq \unit[1.7\cdot 10^{-6}]{A/V}$} result. For a comparison of these values with literature data of other magnetoelectric crystals it is more convenient to use dimensionless linear magentoelectric coefficients $\alpha^{\mathrm{dl}}_{ij}$  as introduced by \cite{Hehl2008,Rivera2009}. Here, $\alpha^{\mathrm{dl}}_{ca}\simeq 4.6\cdot 10^{-4}$ and $\alpha^{\mathrm{dl}}_{cc}\simeq 6.4\cdot 10^{-4}$. From these values the dimensionless rationalized Gaussian coefficients $\alpha^{\mathrm{rG}}_{ij}$ can easily be determined as follows~\cite{Rivera2009}:
\[
\alpha^{\mathrm{rG}}_{ij}=\frac{1}{4\pi}\sqrt{\mu_0/\epsilon_0} \;\alpha^{B}_{ij}=\frac{1}{4\pi}\alpha^{\mathrm{dl}}_{ij}\simeq 
\left\lbrace\begin{array}{ll}
3.7 \cdot 10^{-5},  & \textnormal{for }ij=ca\\
5.0 \cdot 10^{-5},  & \textnormal{for }ij=cc 
\end{array}\right.
\]
These values lie within the typical range of $\alpha^{\mathrm{rG}}$ of $10^{-6}$ to $10^{-2}$, as can be seen from a compilation of data given in~\cite{IntTab}.
Note, that the non-ferroelectric, antiferromagnetic phase of (NH$_4$)$_2$[FeCl$_5$(H$_2$O)] shows no linear magnetoelectric effect, in contrast to K$_2$[FeCl$_5$(H$_2$O)], which, according to our preliminary investigations, is linear magnetoelectric in its antiferromagnetically ordered phase. Consequently, the non-ferroelectric, antiferromagnetic phase of (NH$_4$)$_2$[FeCl$_5$(H$_2$O)] has a symmetry different from the symmetry $m'm'm'$ of the antiferromagnetic phase of K$_2$[FeCl$_5$(H$_2$O)].

\section{Conclusion}

The results of the present investigations of a series of thermodynamic properties reveal that (NH$_4$)$_2$[FeCl$_5$(H$_2$O)] is a new multiferroic material. Multiferroicity in this compound arises at $\sim$\,\unit[6.9]{K}, with preceeding AFM ordering at $\sim$\,\unit[7.3]{K}. While for low applied magnetic fields an electric polarization of about \unit[3]{$\mu$C/m$^{2}$} is lying within the $ab$ plane, it can be significantly enhanced and rotated to $\bi{c}$ for magnetic fields above about \unit[5]{T}. The rather complex temperature versus magnetic field phase diagrams of (NH$_4$)$_2$[FeCl$_5$(H$_2$O)] show several multiferroic/magnetoelectric phases, that differ in orientation and magnitude of the electric polarization and also in the orientation of the magnetic moments of iron. In order to characterize the different multiferroic/magnetoelectric phases of (NH$_4$)$_2$[FeCl$_5$(H$_2$O)] on the microscopic scale, and thus to enable an understanding of the underlying mechanism of multiferroicity of this compound, a detailed determination of the magnetic structures by means of neutron scattering is essential as a future investigation. However, because of the high content of hydrogen atoms (10 H per formula unit) this will require the use of the analogous deuterium compound (ND$_4$)$_2$[FeCl$_5$(D$_2$O)]. Of course, a substantial influence of the isotope exchange on the relevant crystal properties is expected and will need a detailed investigation.

\clearpage

\section*{Acknowledgements}
We thank T.~Bardenheuer and J.~Brand for assistance during the X-ray powder measurements and S.~Heijligen for measurements of the magnetic susceptibility.
This work was supported by the Deutsche Forschungsgemeinschaft via SFB 608 and via the Center of Excellence {\it Quantum Matter and Materials} (QM$^{2}$) at the University of Cologne.

\end{document}